# *Integrating design synthesis and assembly of structured objects in a visual design language*

OMID BANYASAD, PHILIP T. COX
*Dalhousie University, Halifax, Nova Scotia, Canada B3H 1W5*
(*email:* `{banyasad, pcox}@cs.dal.ca`)

**Abstract**

Computer Aided Design systems provide tools for building and manipulating models of solid objects. Some also provide access to programming languages so that parametrised designs can be expressed. There is a sharp distinction, therefore, between building models, a concrete graphical editing activity, and programming, an abstract, textual, algorithm-construction activity. The recently proposed Language for Structured Design (LSD) was motivated by a desire to combine the design and programming activities in one language. LSD achieves this by extending a visual logic programming language to incorporate the notions of solids and operations on solids.

Here we investigate another aspect of the LSD approach; namely, that by using visual logic programming as the engine to drive the parametrised assembly of objects, we also gain the powerful symbolic problem-solving capability that is the forté of logic programming languages. This allows the designer/programmer to work at a higher level, giving declarative specifications of a design in order to obtain the design descriptions. Hence LSD integrates problem solving, design synthesis, and prototype assembly in a single homogeneous programming/design environment. We demonstrate this specification-to-final-assembly capability using the masterkeying problem for designing systems of locks and keys.

## 1. Introduction

The many definitions of the term "design" do not agree in detail. A notion common to them all, however, is that design is a purposeful activity, oriented towards producing some *descriptions* from a set of *specifications* that describe the *function* that the designed artifact is to achieve. Descriptions resulting from design activity contain components of the design and their relationships, and represent the designed artifact in a form that facilitates manufacture, assembly, or construction (Gero, 1990). *Design compilation* is the automatic generation of the design descriptions from specifications.

Design is an essential part of many creative activities. Notably, it is an integral part of engineering practice. Hence, good engineering requires good design. Good design, in turn, calls for appropriate design tools and methodologies (Waldron, 1999).

Electrical and mechanical engineering are two examples of problem domains in which design tools play a pivotal role, to the extent that designing today's state-of-the-art Very Large Scale Integrated (VLSI) circuits and complex mechanical products without employing the right tools seems unimaginable if not impossible.

Although there are many design tools for both electrical and mechanical engineering, Whitney notes that design methods and Computer-Aided Design (CAD) tools for mechanical design are not as mature as those for VLSI. Whitney (1996) associates this gap with the fundamental differences between the two domains despite their superficial similarities, and concludes that design tools and methodologies for mechanical engineering cannot be like those for VLSI design because of these differences.

There are, however, more optimistic views of design tools for mechanical systems. For example, while acknowledging many points raised by Whitney, Antonsson (1997) notes that Whitney's comparison focuses on a very specific subcategory of mechanical design, namely high-powered electro-mechanical design, and concludes that automatic mechanical design can be achieved if researchers focus on such issues as languages for specifying mechanical function, and look for sub-domains of mechanical systems in which logical homologues or homomorphisms (or other mappings) between system requirements and embodiments can be established.

Despite many achievements in developing design tools for VLSI and mechanical design (more in the former domain), today's CAD systems either provide poor facilities for visual design (VHDL, 1988), or provide visual design facilities without well integrated programming capabilities (Autodesk, 1992). The primary reason for including programming capabilities in design environments is to allow for the representation of *parametrised* designs: that is, designs that represent families of artifacts, rather than single objects. We believe that programming capabilities, however, are also necessary for addressing the shortcomings of mechanical design systems, mentioned above.

Visual software tools for designing solid objects, such as AutoCAD (Autodesk, 1992) and MicroStation (Bentley Systems, 2001), evolving as they have from graphical drawing packages, provide tools for solid modeling using concrete visual representations of the components of the design artifacts. Typically, such systems also provide textual programming languages to allow for design parametrisation. For example, AutoCAD supplies AutoLISP for programming, as well as allowing connections to modules written in other textual languages. In such environments, graphical drafting and textual programming are two separate activities performed with very dissimilar tools.

In contrast, the tools most commonly used for designing VLSI devices are purely textual languages, akin to ordinary programming languages, and providing no visual design capability. This is a consequence of the fundamental differences between the design of solid objects, where the designer concentrates on physical components and relationships, and VLSI design, where the designer's concern is rather more abstract, namely, implementing Boolean functions (VHDL, 1988).

In another thread, *design languages* for formally describing the constituent components and structure of objects, have been devised and extensively studied. Some of these languages have their genesis in robot programming, and are aimed at representing plans that a robot can use to assemble structured objects out of parts, for example (Popplestone, Ambler & Bellos, 1980, Rocheleau & Lee, 1987 and Gottschlich & Kak 1994). Some arise from similar assembly considerations, but are motivated by computer-aided manu-facturing (Nackman et al., 1986). Others result from a desire to provide formal and general underpinnings for geometric modelling in CAD systems (Snyder, 1992, Heisserman & Woodbury, 1993 and Paoluzzi, Pascucci & Vicentino, 1995). The language PLASM, for example, is a functional language in which geometric objects are defined as functions (Paoluzzi, Pascucci & Vicentino, 1995), and more complex objects are created by combining functions using appropriate functional operators. A PLASM function can be evaluated to produce representation of the object it describes. This representation can then be exported to other formats and used by other application to produce a 3D rendering. Clearly, this model for solids provides for parametrisation, since a function generates a variety of different objects depending on the values provided for its formal parameters. Also, parameters can themselves be functions representing solids, so the notion of *operations* on solids is also captured.

Design languages for solid objects are analogous to languages such as VHDL for VLSI design in that they provide very powerful programming capability in a *textual* language, but lack the direct visual manipulation provided by CAD environments. The authors of PLASM acknowledge the necessity for visual representations, saying *"Our long term goal is to develop a visual language shell around a PLaSM nucleus. Thus, it will become possible to build personalized graphics interfaces where both the designer/user and the PLaSM application programmer will be comfortably accommodated"* (Paoluzzi, Pascucci & Vicentino, 1995, p300). Their intention, therefore, was not to build a single visual language, but a system within which different visual languages could be built for the designer and the programmer. To our knowledge, no such visual language shell for PLASM has been developed.

In an effort to remove the sharp distinction between designing and the programming required to achieve parametrisation, mentioned above, and to represent both activities visually, the visual design language LSD (Language for Structured Design) was proposed by Cox and Smedley (1998). LSD is based on a visual logic programming language, Lograph (Cox & Pietrzykowski, 1985), extended by the introduction of "explicit components", a logical manifestation of solid objects, and a new execution rule to combine them.

The original LSD, as a simple exploration of the notion of design in a visual language, had just one operation on solids. However, the selection of operations required in a design language depends on the domain of application, so LSD clearly needed to be extensible. As a basis for extensibility, a very general notion of a design space was required, providing just enough detail to capture the essence of solids in logic. To meet this need, a formal model for characterising objects in a design space was proposed by Cox and Smedley (2000), where a design space consists of multidimensional space (normally 3-D) augmented with extra dimensions representing properties. This formalisation also defines the concept of an operation on solids, and selective interfaces that allow solids to declare what operations can be applied to them. Based on this model, the required generalisation of LSD was made. It is important to note that this formalisation characterises solids and operations only to the extent required by LSD, so does not address any of the practical details of solid modeling. Nevertheless, we will refer to it as the "solid modeler" in the following.

As a visual language for describing parametrised designs, LSD has similar characteristics to the visual language shell proposed by the designers of PLASM. Specifically, it is a visual language shell that can be customised by the addition of domain-specific components, and operations, and accommodates both designer and programmer (who could be the same person).

Interestingly, the work on design languages in the CAD community has concentrated on the ability of these languages to describe designs, without examining their potential for solving design problems in order to generate design descriptions. This may be a consequence of the fact that solving design problems frequently involves searching large spaces, a task directly supported by few languages other than logic-based ones. For example, the automated house-design system reported by Rau-Chaplin, MacKay-Lyons and Spierenburg (1996) is implemented in Prolog, and applies rules to generate and filter very large spaces of house designs based on criteria specified by the user.

Although the original impetus behind LSD was to provide a visual language for designing parametrised objects, the fact that it inherits logic-programming problem-solving and search capabilities from Lograph enables LSD to also address *design specification*. The user of an LSD-based CAD system could, therefore, construct a visual logic program that implements a solution to a design problem. Executing a query that specifies an instance of this problem would generate a design description and assemble the resulting artifact in one continuous, observable process. This possibility, initially proposed by Banyasad and Cox (2002b), is explored below via a detailed example using the masterkeying problem.

CAD systems are inherently visual. The designer "debugs" by observing the designed artifact from different angles, in different representations (e.g wireframe), etc., and refining it to remove errors. When programming for parametrising or problem solving is added, interpretive execution must be included in the debugging process. In LSD, execution rules can be animated as graph transformations. This raises certain implementation issues, which although not central to the main theme of this paper, are important if a practical LSD-based CAD system is to be developed. We will discuss these briefly in Section 4.

## 2. Language for Structured Design

LSD is an extension of Lograph, a general-purpose, visual, logic programming language, and is intended for designing structures rather than general programming; hence the names of LSD entities have been chosen to be suggestive of their roles in the design world.

LSD provides capabilities both for programming and for designing parametrised components of complex objects in a homogenous environment. A component designed in LSD represents a family of solids, individual members of which can be realised by giving specific values to design parameters. For example, a particular kind of parallelogram may be defined in a 2-dimensional design world by a fixed height and variable base lengths. If the parallelogram is incorporated in a larger design, its bases may become constrained to some specific values when the parallelogram is fused to other shapes. Parametrising designs using the programming features of LSD in a descriptive fashion, and manipulating designs in a solid world are features of LSD that will be exploited in the following sections.

Since LSD is central to our presentation, we will briefly and informally introduce it using the example of designing keys. Because LSD is a logic programming language, it has some similarities with Prolog (Kowalski, 1979). We will rely on those similarities to simplify our description of LSD. Formal and comprehensive discussions of Lograph and LSD can be found elsewhere (Cox & Pietrzykowski, 1985, Cox & Smedley, 1998).

A key can be described as consisting of four types of components, a *handle*, a number of *bits*, a number of *levellers*, and a *tip*. Figure 1 depicts several such components, known as *explicit components* (*e-components*) in LSD. Each explicit component may have open edges indicated by arrows, along which it can be bonded to other solids. The solid lies to the right side of the arrow, when observed from the point of view of someone standing at the tail of the arrow and facing in the direction indicated by the arrow. We will omit these arrows in later diagrams.

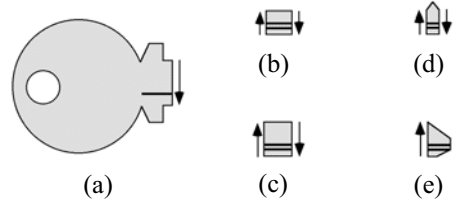

Fig. 1: Components of a key solid with two possible bit cuttings.

The handle and tip, Figure 1(a) and (e), have no logical significance in the functionality of a key. A bit is a solid of a fixed width and height with open edges at its sides. In our example, as illustrated in Figure 1(b) and (c), a bit comes in two possible sizes, differing only in height. A leveller is a solid of fixed width as in Figure 1(d). The heights of the sides of a leveller can be independently varied. When the side of a leveller is fused to another solid, the height of the leveller at the binding side is constrained to the height of the other solid. Once both sides of a leveller are bonded to other solids, the leveller will create a ramp between the tops of those solids. The solid bonded to the left of a leveller may be a bit or a handle. The solid bonded to the right of a leveller may be a bit or a tip. The height of the tip at its left side is variable and will be set to the height of the solid to which it is bonded.

Explicit components in an LSD design represent solids that will be combined into larger structures when the design is executed. LSD designs also contain other objects representing the traditional data structures and operations of programming languages. Since LSD is a logic programming language, its data structures are functional expressions, like terms in Prolog, but represented pictorially as we shall now explain.

In Prolog, functors are used to create terms which can be viewed as data structures. In LSD *function cells* are used for a similar purpose. A function cell consists of a name, a *root terminal* and a list of terminals of length $n \geq 0$ called the *arity* of the cell. Figure 2 illustrates a function cell named • with arity 2. The small circles in Figure 2 are *simple terminals*. The terminal on the peak of the curved face of the function cell is its root terminal. A function cell can have two possible orientations  or . Regardless of the orientation, the terminals of the function cell are ordered from left to right. A function cell with arity 0, also called a *constant*, has the simpler representation, <u><name></u> or <u><name></u>, where <name> is the name of the cell.

A list is a data structure which is either empty or non-empty. In Prolog, an empty list is represented by the Prolog atom, []. A non-empty list consists of a head and a tail combined with the special • functor. The head can be any Prolog term and the tail must be a list. Pro-log provides a special notation for lists; namely, the string [item-1, item-2,....., item-n] represents the term •(item-1,•(item-2,…•(item-n,[])...)). Similarly, in LSD one can use function cells named • to create lists, and the constant 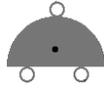 to signify the empty list. The root terminal of a • cell represents the list, while its first and second terminals correspond to the head and tail of the list respectively. For example, the graph in Figure 3 corresponds to the Prolog term •(1,•(2,•(1,•(2,[])))) and represents the list[1,2,1,2]. The line segments that connect terminals are called *wires*. In LSD, as in Prolog, a list can be abbreviated. For example, the list in Figure 3 can be denoted $\overline{{}_{[1,2,1,2]}^{\circ}}$ .

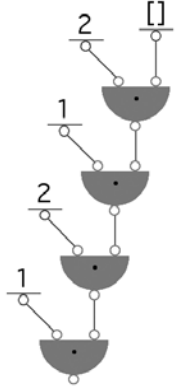

Fig. 2: A function cell

Fig. 3: A list in LSD

An LSD *program* is a set of designs. A *design* is a set of cases which define the structure of an artifact. A *case* consists of a name, a head and a body. The *head* of a case is an ordered list of terminals of length $n$ for some integer $n \geq 0$ called the *arity* of the case. In the pictorial representation of a case, the head is a rounded rectangle with a clockwise-pointing arrow on it, called the *origin*, and the terminals arranged around the perimeter starting from the origin. A terminal is either a simple terminal represented by a small circle, or an *edge terminal*, represented by ▪▪▪▪▪▪▪▪ .

The *body* of a case is a network of cells interconnected by wires and bonds. A *bond* connects two edges, where an *edge* is either an edge terminal or an open edge. A *cell* is either a function cell, or a component.

Figure 4 depicts a program consisting of two *designs*, **Partial-Key** and **Key**. **Partial-Key** expects to receive a list of integers on its single simple head terminal, each corresponding to a bit of a key, and recursively describes a partial key as the component obtained by bonding a bit corresponding to the head of its input list to a partial key corresponding to the tail of its input list. The handle in the single case of the design **Key**, and the tip in the non-recursive case of **Partial-Key**, are *explicit-components* (e-components). An *implicit component* (i-component), consisting of a name and a list of terminals, is pictorially represented by a rounded rectangle with a clockwise-pointing arrow called the *origin* on its perimeter, and the terminals arranged around the perimeter starting from the origin. For example, the round rectangles named **Partial-Key** are implicit components representing an invocation of the **Partial-Key** design. The grey stripe connecting the open edge of the handle and i-component **Partial-Key** is a bond. A *bond* connecting two e-components fuses those two components at their corresponding open edges during execution.

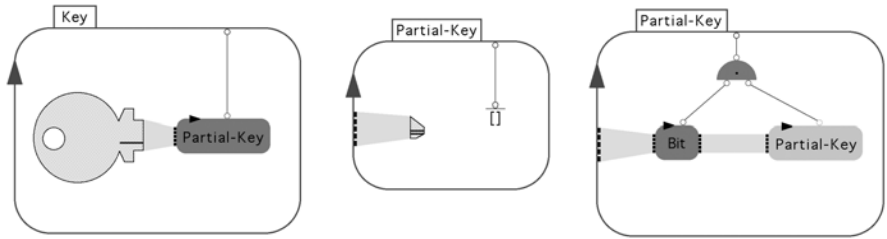

Fig. 4: The LSD designs **Key** (single case) and **Partial-Key** (two cases)

Each design and i-component has a *signature* consisting of its name together with a list of the types of its terminals. For example, the design **Partial-Key** and the i-components **Partial-Key** in Figure 4 all have signature (**Partial-Key**, (simple, edge)).

As the reader may have noticed, the two i-components in the recursive case of design **Partial-Key** are painted in two different shades of the same colour. This is a practical issue that will be clarified later.

The implicit component **Bit** in the recursive case of **Partial-Key** is an invocation of a design **Bit** shown in Figure 5, defined by two cases corresponding to the two sizes of bit.

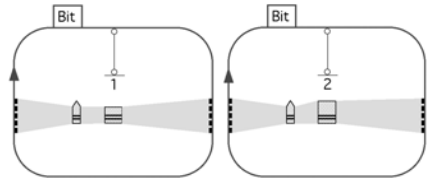

Fig. 5: Two cases of the design **Bit**.

The process of building a component according to the designs in a program is called *assembly*. Assembly transforms a *design specification*, a network of function cells and components, using four execution rules: *replacement*, *merge*, *deletion* and *bonding*. For example, Figure 6(a) depicts a design specification for a key with four bits of sizes 1, 2, 1 and 2. The specification in Figure 6(b), parts of which have been omitted for compactness, is obtained by applying the replacement rule to the one in Figure 6(a). In this transformation, the i-component **Key** is replaced with a copy of the body of the case of the design **Key**. During replacement, connections (bonds or wires) are established by matching terminals of the replaced i-component with corresponding terminals in the head of the case.

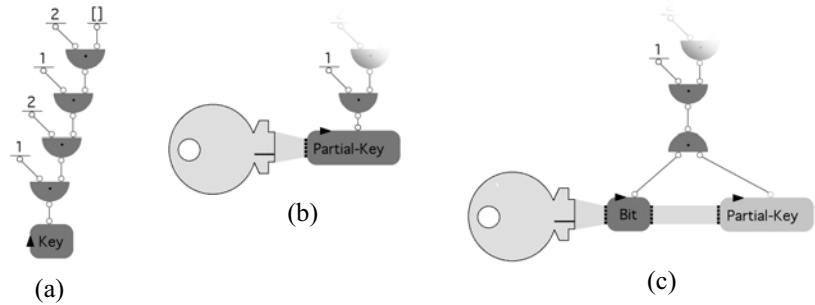

Fig. 6: A key specification, and the replacement rule.

Applying replacement to the i-component **Partial-Key** in Figure 6(b) using the recursive case of design **Partial-Key** results in the specification in Figure 6(c), which contains two functions with the same name and arity connected by their root terminals. These are subject to the merge rule which merges the two cells into one, creating new connections by identifying corresponding terminals of the two cells, as shown in Figure 7(a).

Next, the specification in Figure 7(a) is transformed to that in Figure 7(b) by the deletion rule, which removes function cells with unconnected root terminals. The i-component

**Bit** in Figure 7(b) is replaced with the case of design **Bit** on the left of Figure 5 resulting in the specification shown in Figure 7(c). One application of merge and one of deletion removes the two constants named 1 connected producing the specification in Figure 8(a).

One of the consequences of the last replacement is that the specification now contains bonds which are incident on open edges rather than edge terminals. Such bonds can be executed, joining together the e-components at their ends. In our example, executing the two executable bonds produces the specification in Figure 8(b), containing a new e-component, a partially defined key with one bit and no tip.

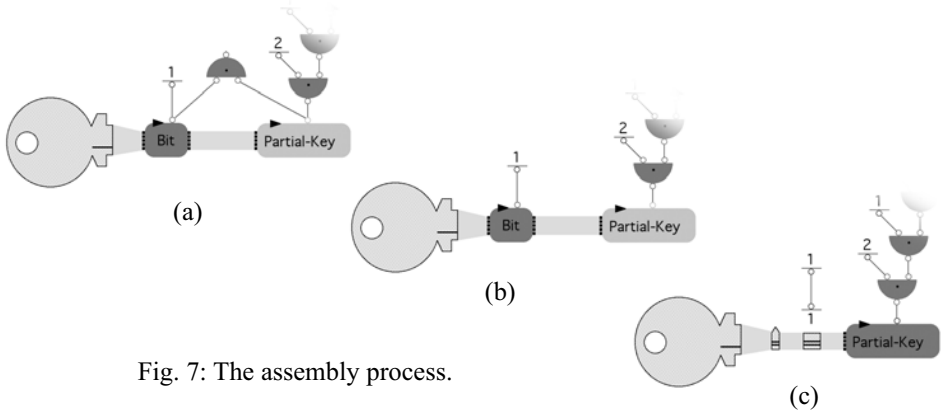

Fig. 7: The assembly process.

Continuing execution will transform Figure 8(b) into the e-component in Figure 8(c) by adding more bits until all items in the list of the original specification are consumed. Assembly stops when a specification is produced that cannot be further transformed, as depicted in Figure 8(c).

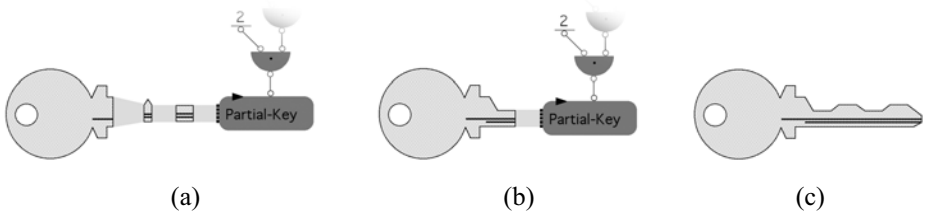

Fig. 8: Key assembly process.

The reader has no doubt noticed that by choosing the "correct" case of **Bit** in the replacement that transformed Figure 7(b) into Figure 7(c), we guaranteed that the merge and deletion rules would apply to obtain Figure 8(a). If we had made the wrong choice, the graph would have included the constants ⊤/1 and ⊥/2 connected by their roots. All subsequent graphs in the sequence would also contain this subgraph since none of the execution rules apply to it. Consequently, although the final, untransformable specification would contain a key, it would also contain this unremovable connected constants, and perhaps others resulting from "incorrect" choices. Such assemblies are considered to be unsuccessful.

Having given a general flavour of LSD, we will now focus on its problem-solving capabilities.

### 3. Problem Solving and Design

As we noted earlier, LSD extends Lograph by including solids and operations on solids. However, many of the computations required in the synthesis phase of design may be purely symbolic or arithmetic. Such computations can be performed by the Lograph subset of LSD, which we identify as follows. In LSD, a *definition* is a design with the property that every i-component occurring in a case of the design is a literal. A *literal* is an i-component with the same signature as a definition. Hence a literal is an i-component that does not introduce any e-components, either directly or recursively. A Lograph program is one that consists of definitions. Note that "successful" execution of a specification that consists only of function cells and literals results in the empty graph.

In this section, we explain how the problem-solving capabilities provided by the Lograph subset, together with the design specification aspects of LSD can provide a homogeneous programming/design environment for solving some structured artifact design problems. We will use *masterkeying* as an example to illustrate the process. A comprehensive discussion of the masterkeying problem and the analysis of its complexity can be found in (Espelage & Wanke, 2000) on which we have based our definitions.

#### *3.1 Masterkeying*

*Masterkeying* is allowing several different keys to operate one lock. It is used in a systematic way to regulate access to areas according to key holders' privileges. In a master key system, there is always one master key with the highest access privilege that can operate every lock in the system, and there may be sub-masters of various levels that can open subsets of the set of locks. A change key is one that can operate only one lock. A more precise definition of the masterkeying problem follows.

A *key* is a solid that can open a *lock*. A lock is also a solid consisting of a *body*, and a *plug*. The body has a set of *chambers*. Inside every chamber is a *spring* and a *pin*. Every spring is attached to the end of the chamber at one end and the pin at the other. Every pin is cut through at some level. The end of a pin connected to the spring is called the *driver pin* while the other end is called the *key pin*. A plug is a rotating cylinder contained inside the body. A plug has a *key way* and a set of *holes* through which the pins can slide. When a key is inserted in the keyway, the pins are pushed towards the body a distance depending on the shape of the key. A key can *open* a lock when the key pushes all the pins in such a way that the cut in each pin is aligned with the shear line between the body and the plug. A wrong key will leave at least one pin across the shear line, preventing the plug from turning. Figure 9 illustrates the side view of a lock and one of its keys.

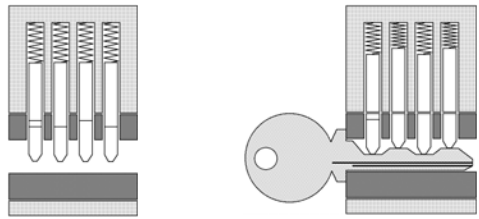

Fig. 9: A side view of a lock and one of its keys.

In a master key system, a lock has at least one pin with more than one cut, allowing more than one key to open it. The two parts of a pin at the body side and the key side are still called driver and key pin respectively, and the parts in between are called *master pins* or *spacers*.

A *locking system* is a tuple $\mathcal{L} = (s_1, s_2,..., s_k)$ where $k \geq 0$ is the number of chambers, and for each $i$ ($1 \leq i \leq k$), $s_i \geq 1$ is the maximum number of possible cut levels for the pin in the $i^{th}$ chamber. A *key in* $\mathcal{L}$ is described by a *bitting vector,* which is a k-tuple $(b_1, b_2,..., b_k)$ such that $1 \leq b_i \leq s_i$ for each i ($1 \leq i \leq k$). A *lock in* $\mathcal{L}$ is described by a *bitting array* which is a k-tuple $(C_1, C_2,..., C_k)$ such that $C_i$ is a subset of $\{1,2,...,s_i\}$ for each i ($1 \leq i \leq k$). For example, the key and lock in Figure 9 are described by the bitting vector (1,2,1,2) and bitting array $\{\{1,2\},\{2\},\{1\},\{2\}\}$.

A key with bitting vector $(b_1, b_2,..., b_k)$ *opens a lock* with bitting array $(C_1, C_2,..., C_k)$ iff $b_i \in C_i$ for all $i$ such that $1 \leq i \leq k$.

Given a list of *n* keys such that the bitting vector of the $j^{th}$ key in the list is $(b_{j1}, b_{j2}, ... , b_{jk})$, a lock that is opened by every key in the list is defined by the *induced bitting array* $\mathcal{A} = (C_1, C_2, ... , C_k)$ where $C_i = \{b_{ji} \mid 1 \leq j \leq n \}$ for all $i$ between 1 and $k$. For example, an induced bitting array for keys defined by bitting vectors (1,2,2,1), (2,2,2,1), and (1,1,2,1) is $\{\{1,2\},\{1,2\},\{2\},\{1\}\}$.

A *key-lock matrix* $\mathcal{X} =(x_{ij})$ is an $n \times m$ matrix where *n* and *m* are, respectively, the numbers of keys and locks in the system, $x_{ij} = 1$ if key $i$ can open lock $j$ and $x_{ij} = 0$ otherwise. Table 1 shows a key-lock matrix for a system with three keys and two locks.

An *implementation* of a key-lock matrix $\mathcal{X}$ is a list $(\mathcal{B}_1,...,\mathcal{B}_n)$ of $n$ bitting vectors describing $n$ keys, and a list of $(\mathcal{A}_1,...,\mathcal{A}_m)$ of $m$ bitting arrays describing $m$ locks such that key $i$ can open lock $j$ if and only if $x_{ij}$ of $\mathcal{X}$ is 1. For example, the list of bitting vectors (1,2,1,2), (2,2,1,2), and (1,2,2,2) together with the list of bitting arrays $\{\{1,2\},\{2\},\{1\},\{2\}\}$ and $\{\{1\},\{2\},\{1,2\},\{2\}\}$ is a possible implementation of the key-lock matrix of Table 1.

The *masterkeying problem* is as follows. Given a key-lock matrix $\mathcal{X}$ and a locking system $\mathcal{L}$, find an implementation of $\mathcal{X}$ such that the bitting vectors and arrays of the implementation describe keys and locks in $\mathcal{L}$.

Espelage and Wanke (2000) show that a masterkeying problem for a given number of pins where every pin has exactly two cut levels is NP-complete. There are, however, a number of heuristic methods for solving this problem. We assume a maximum number of cut levels for each pin and choose a bitting vector for the top master key as the starting point. Then we derive the bitting vectors of the lower level keys in a systematic way. We employ this simple heuristic method in an LSD program to find an implementation of the key-lock matrix of Table 1.

|  | Lock 1 | Lock 2 |
|---|---|---|
| **Master key** | 1 | 1 |
| **Change key 1** | 1 | 0 |
| **Change key 2** | 0 | 1 |

Table 1: A key-lock matrix

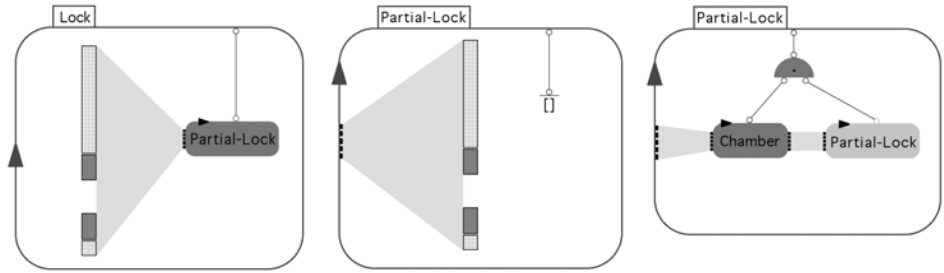

Fig. 10: Cases of **Lock** and **Partial-Lock**.

### *3.2 Keys and Locks in LSD*

In Section 2 we showed how a key solid can be expressed in LSD terms. We use a similar approach to define a lock, as illustrated in Figure 10. **Lock** is a design with a single case that defines a lock from a bitting array, represented as a list of rows. The e-component occurring in this design is the front shell of the lock. The recursive case of the design **Partial-Lock** detaches the head of the input list and uses it to add an appropriate chamber to the assembled partial lock, then recursively generates the rest of the lock according to the remaining items in the tail of the input list. The e-component in the non-recursive case of **Partial-Lock** is the back shell of the lock.

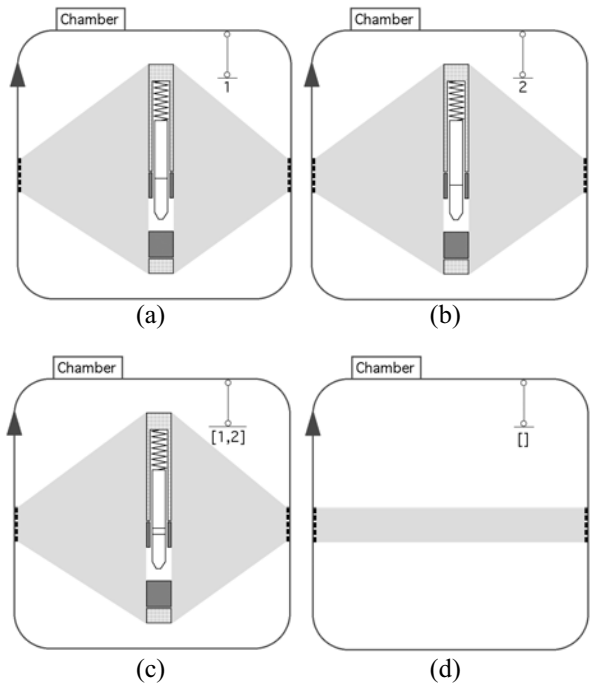

Fig. 11: **Chamber** design.

The implicit component **Chamber** in Figure 10 corresponds to the design depicted in Figure 11. Each of the three cases labelled (a), (b), and (c) corresponds to one out of three possible pin cut configurations. Case (d) in Figure 11 corresponds to a null chamber.

In order to express a masterkeying problem in LSD, we define two additional literals: **Open** and **Member**, the former checks whether a key can open a lock and the latter defines the relation "member of list". Their definitions are shown in Figure 12. Note that in LSD, an unconnected terminal is analogous to an anonymous variable in Prolog.

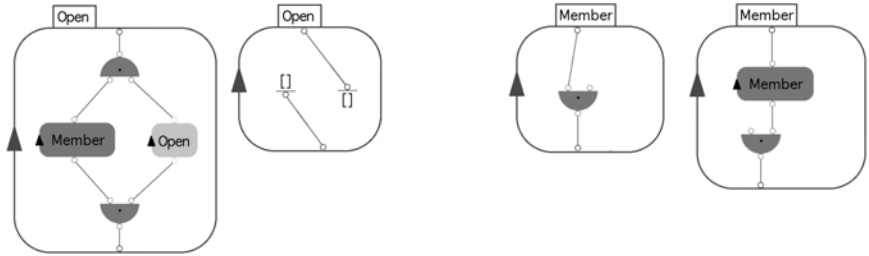

Fig. 12: **Open** and **Member** definitions.

### 3.3 Key-Lock Graph

Now that we have the necessary definitions at the component level, we can construct the system level specification of the problem in Figure 13 which expresses a solution to a masterkeying problem in which there are three keys (a master key and two change keys) and two locks. In this specification, the **Open** literals correspond to the 1's in the key-lock matrix of Table 1, and the *crossed* **Open** literals correspond to the 0's. Note that crossed lines on a literal denote negation.

The graph in Figure 13 is, therefore, a visualisation of the key-lock matrix of the masterkeying problem to be solved, assuming a bitting vector [1,2,1,2] for the master key. Execution will produce bitting vectors for the two change keys and induced bitting arrays for the two locks, which serve as parameters to the corresponding **Key** and **Lock** i-components, from which corresponding e-components are assembled. Figure 14 shows the final result after execution of the specification is complete.

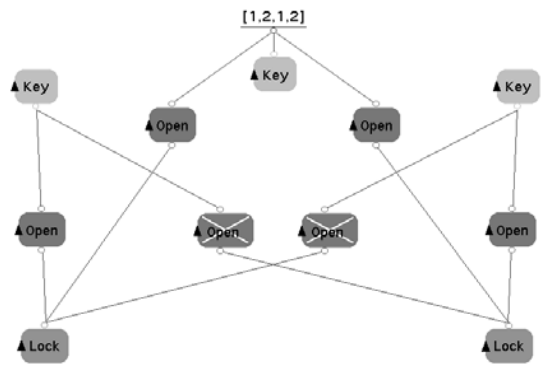

Fig. 13: Masterkeying problem in LSD.

As we mentioned earlier, the different shades of i-components in a design or specification is a practical issue, relating to the implementation of LSD. The language, as we have described it, is non-deterministic, so some ordering mechanisms are necessary to ensure complete and efficient traversal of the search space, as in Prolog. The different shades in which i-components are painted indicate the order in which the components will be executed using the replacement rule, where darker components are executed first. When a replacement rule introduces a new set of i-components, the new components are painted in shades darker than existing components, giving them higher priority. After every replacement rule, all i-components in a specification are recoloured to comply with the ordering mechanism. The programmer imposes this ordering when building designs and specifications (Banyasad & Cox, 2001).

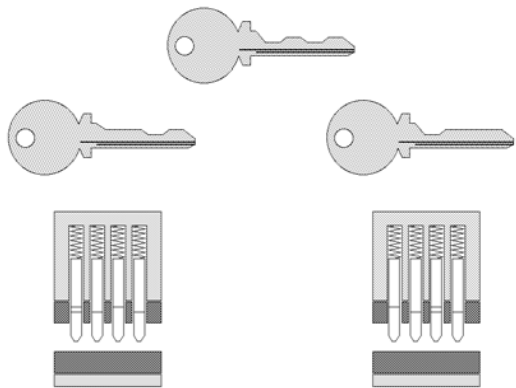

Fig. 14: Query result.

Note that all **Open** literals in the specification in Figure 13 should be executed before any **Key** or **Lock** i-component since they compute the parameters for assembly. This divides the execution of the specification into two parts, solving the design problem, followed by constructing the keys and locks according to the computed parameters. From the designer's point of view, however, the diagram in Figure 13 simply combines problem-solving and construction in a single declarative specification. This homogenous expression of problem description and construction procedure relieves the designer from having to consider issues surrounding the execution order of problem solving and assembly. Note, however, that as is the case in other logic programming languages, a good choice of execution order may well result in the assembly process being faster and more efficient, and may also affect termination. Given the standard order of the phases of the design cycle, it should be quite natural for a designer creating such specifications to impose the correct ordering on the components of a design.

No doubt the reader has noticed that the masterkeying design process has been decoupled into two phases, the system level, specified by the query of Figure 13, and the component level expressed by the definitions of **Key** and **Lock**. This creates a clean separation between system-level design and component-level design.

### 4. Animating the assembly process

As noted earlier, LSD is a Visual Programming Language (VPL) and like any other VPL can benefit from an appropriate Visual Programming Environment (VPE). VPEs assist programmers to build, edit, debug and execute programs by incorporating appropriate graphical representations. For example, Prograph CPX is a VPE which provides the VPL Prograph as the programming language (Prograph, 1993).

Animating the execution of LSD programs (assemblies) can provide the user with a smooth visual transformation of a design specification into the final design description which is continuous with respect to time and space. This continuous transformation from one state of the design to the next makes it is easier for the user of LSD to comprehend the semantics of the design in contrast with a discontinuous transformation in which LSD execution rules are treated as atomic operations. Hence, the animation of an assembly can help the user to better understand, edit, and debug LSD programs.

Transformation of the design specification in Figure 13 to the solid products in Figure 14 is achieved through the application of the four LSD execution rules, merge, deletion, replacement, and bonding. An implementation of LSD could inherit the animation of the first three execution rules from its underlying Lograph implementation. For example, our prototype implementation of Lograph, discussed by Banyasad and Cox (2001), animates the deletion of a function cell by first letting go of all the wires connected to the cell's terminals, then fading the cell away. The merge rule is animated by moving one of the function cells over the other, then deleting one of them. The replacement rule is animated as an expansion of the literal being replaced, and a fading in of the cells in the body of the case used in the replacement. The animation of the bonding rule is, however, a more complex issue since it operates on explicit components.

Explicit components are the representatives in LSD of solids, the physical properties of which are maintained by an associated solid modeler and unknown to LSD. Clearly, therefore, the visual representations of e-components, such as the various parts of keys and locks in the examples above, cannot be drawn by LSD. Similarly, the animation of assembly requires knowledge of the physical properties of the associated solids. Hence drawing and animation of e-components must be dealt with by the associated solid modeler. Since solid modeling is pivotal to our discussion on how the assembly process is animated, we will give a brief overview of the formal definition of design spaces and solids proposed by Cox and Smedley (2000).

Solids are modeled in a *design space* which is a multidimensional (usually 3D) space augmented with an arbitrary but fixed, finite number of real-valued *properties*. A *solid* is a function that maps a list of *parameter values* to a set of points in the design space, constituting the volume of the solid. Therefore, each solid in the design space represents a family of solids, each realised by a particular choice for parameter values.

Bonding, used in our examples above, is an example of an operation for creating complex solids from simpler ones. An *operation* in a design space $\mathcal{D}$ is a 3-tuple ($\mathcal{F}$, $\mathcal{L}$, $C$) where for some integer $n > 0$, $\mathcal{F}$ is a function from $\mathcal{D}^n$ to $\mathcal{D}$ which, when applied to $n$ operand solids in $\mathcal{D}$, defines the set of points in the new solid. $\mathcal{L}$ is a list of $n$ selectors, one for each operand. A *selector* **L** is a formula, one of the free variables of which corresponds to the solid to which it is applied. The remaining variables of a selector **L** identify variables of the operand that are required by the operation, and are extracted from the operand solid via the application to the solid operand of a partial function, uniquely determined by **L** and the solid, called an **L**-*interface*. $C$, the *constraint* of the operation, is a formula, the free variables of which are the variables extracted from the operands by the selectors. An operation is *successful* if $C$ is satisfied by the values of the variables extracted from the operand solids by the selectors. For example, let

**Punch** = $(\cup, \{\textbf{centre}(a_1,b_1,c_1,r_1), \textbf{centre}(a_2,b_2,c_2,r_2)\}, \textbf{\textit{C}}(b_1,c_1,r_1,b_2,c_2,r_2))$
where **centre**$(a,b,c,r)$ is a formula with free variables $a,b,c$ and $r$ which is true iff $a$ is a disk centred at $(b,c)$ with radius $r$, and $\textbf{\textit{C}}(b_1,c_1,r_1,b_2,c_2,r_2)$ is the formula $[b_1=b_2 \wedge c_1=c_2 \wedge r_2 \leq r_1]$. Then **Punch** is a binary operation in a 2D design space which applies only to disks. Let $p$ and $q$ be two disk solids defined by the function **Disk**$(b,c,r)=\{(x,y) \mid x,y \in \textbf{R}$ and $(x-b)^2+(y-c)^2 \leq r^2\}$. Applying **Punch** to $p$ and $q$ creates a new solid defined by **Ring**$(b_1,c_1,r_1,b_2,c_2,r_2)=\{(x,y) \mid x,y \in \textbf{R}$ and $(x,y) \in \downarrow(\textbf{Disk}(b_1,c_1,r_1) - \textbf{Disk}(b_2,c_2,r_2))$ and $[b_1=b_2 \wedge c_1=c_2 \wedge r_2 \leq r_1]\}$. Note that after the application of an operation, a point in the resulting solid may have two or more values for some property. The role of $\downarrow$ is to reduce the solid to an empty set in such cases. **Ring** is equivalent to **Ring′**$(b,c,r_1,r_2)=\{(x,y) \mid x,y \in \textbf{R}$ and $r_2^2 < (x-b)^2+(y-c)^2 \leq r_1^2\}$.

Bonding in a 2D design space can be defined by the operation $(\cup, (\textbf{edge}(a,y_1,y_2,y_3,y_4), \textbf{edge}(a,y_5,y_6,y_7,y_8)), \textbf{\textit{C}})$ where **edge**$(a, b, c, d, e)$ is true iff $a$ is any solid, $b, c, d,$ and $e$ are real numbers, every point on or to the right of the directed line segment from $(b, c)$ to $(d, e)$ is in $a$ and every point to the left of the line segment is not in $a$; and $\textbf{\textit{C}}(y_1,y_2,...,y_8)$ is the formula $[(y_1,y_2)=(y_7,y_8) \wedge (y_3,y_4)=(y_5,y_6)]$. Note that **edge** identifies variables that define an open edge of a solid, and **C** constrains the selected open edges of two solids to lie on the same line.

Just as bonding is represented by the grey bands in the diagrams above, operations in general need to have visual representation in LSD. A discussion of this issue can be found in (Cox & Smedley, 2000). We will confine our discussion below to the bonding operation.

When an executable bonding operation is encountered in a query, the LSD execution engine initiates a request for the operation to be performed by passing references to the operand solids to the solid modeler. The solid modeler applies the bonding operation to the solids, and depending on whether or not the operation's constraint is satisfied, success or failure is reported back to the LSD engine. A side effect of the application of the operation to the operand solids is that their geometry is adjusted to satisfy the constraint of the operation. How this is achieved is transparent to LSD and depends on how the associated solid modeler works. In the following we will briefly explain how this might be done.

If $x$ and $y$ are two lists, by $x \cdot y$ we denote the concatenation of $x$ and $y$. If A and B are sets, we denote by $(A \rightarrow B)$ the set of all continuous functions from A to B.

*Definition 1*

Let $t_s, t_e \in \textbf{R}$ and $t_s < t_e$. An *n-ary animation* over the interval $[t_s, t_e]$ is a function $\Delta : \textbf{R}^n \times \textbf{R}^n \rightarrow ([t_s, t_e] \rightarrow \textbf{R}^n)$ such that for all $x$ and $y$ in $\textbf{R}^n$, $\Delta(x \cdot y)(t_s)=x$, and $\Delta(x \cdot y)(t_e)=y$. The interval $[t_s, t_e]$ is called the *duration* of the animation.

*Definition 2*

Let $\Phi_1, \Phi_2, ..., \Phi_m$ be $m$ solids in a design space $\mathcal{D}$ where $\Phi_i$ is a function of $n_i$ variables ($1 \leq i \leq m$). Let $\Pi = (\mathcal{F}, \mathcal{L}, \textbf{\textit{C}})$ be an $m$-ary operation in $\mathcal{D}$. Let $\phi_i$ be the interface correspond-ing to $\Phi_i$ and the $i^{\text{th}}$ selector in $\mathcal{L}$ $(1 \leq i \leq m)$, and $\Delta$ be a *k-ary* animation over the interval $[t_s, t_e]$, where $k = \sum_{i=1}^{m} n_i$. If $z \in \mathbf{R}^k$ denote by $z_1$ the first $n_1$ elements of $z$, denote by $z_2$ the next $n_2$ elements of $z$ and so forth. Then $\Delta$ is an *animation* of the application of $\Pi$ to $\Phi_1$, $\Phi_2,..., \Phi_m$ iff there exists $x$ and $y$ in $\mathbf{R}^k$ such that for all $i$ $(1 \leq i \leq m)$ and for all $t$ $(t_s \leq t \leq t_e)$, $\Phi_i(\Delta(x \cdot y)(t)_i) \neq \varnothing$ and $\phi_1(y_1) \cdot ... \cdot \phi_m(y_m)$ satisfies $C$.

The intuition behind the above definition is that an operation's animation can be characterised by how the variables of the operand solids assume values in the period of time over which the operation is animated.

*Example*

Suppose two square solids each have an open edge connected by a bond as shown in Figure 15. Suppose also that the size, orientation and position of each square is unconstrained, and that there are no properties. The variables required from any solid for the bonding operation, are the four variables defining the ends of the open edge to which the bond is attached, shown in the figure. Let $\Delta$ be the function from $\mathbf{R}^8 \times \mathbf{R}^8$ to $([t_s, t_e] \rightarrow \mathbf{R}^8)$ such that for $x, y \in$ $\mathbf{R}^8$, and $t_s \leq t \leq t_e$, $\Delta(x \cdot y)(t) = (f^1_{x \cdot y}(t), f^2_{x \cdot y}(t),..., f^8_{x \cdot y}(t))$ where $f^i_{x \cdot y}(t) = x_i + (t-t_s)(y_i-x_i)/(t_e-t_s)$ for $1 \leq i \leq 8$. Then $\Delta$ is an animation of the application of the bonding operation to the two squares since we can choose values for $x$ and $y$ to satisfy the conditions. For example, choose $x$ in such a way that the squares are sized, positioned and oriented approximately as shown in the figure, and let $y = ((x_1+x_7)/2, (x_2+x_8)/2, (x_3+x_5)/2, (x_4+x_6)/2, (x_3+x_5)/2, (x_4+x_6)/2, (x_1+x_7)/2, (x_2+x_8)/2)$. Then $\Delta$ describes an animation in which the corresponding corners $(x_1, x_2)$ and $(x_7, x_8)$ of the two squares slide toward each other in a straight line, finally coinciding at the point marked **a** in the figure, while the points $(x_3, x_4)$ and $(x_5, x_6)$ meet each other at the point marked **b**. Clearly $y$ satisfies the constraint of the bonding operation, and neither square reduces to an empty set of points at any time.

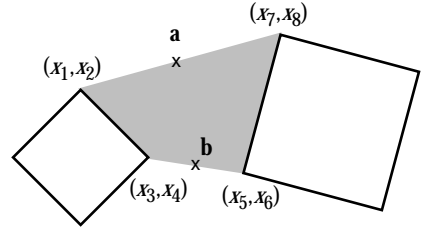

Fig. 15: Bonding animation.

In the above example, the interpolation of the variables of the two operands from their values before the application of the operation to their corresponding values after the successful execution of the operation, is linear with respect to time. However, this need not be the case. For instance, we can rewrite the $f^i$ functions to give the animation a non-linear behaviour. Note that the animation function does not determine which edge is moved to align to the other edge. For example, the left square might be anchored in which case the right square will move towards it so that the open edges match, or the right square might be anchored so that the left square will move. The animation function supports both situations. If neither of the edges are anchored, then depending on how the solid modeler solves and assigns values to the free variables, the edges will move to align accordingly.

The animation of the bonding operation in the above example attaches two edges of the operands in a fashion that would seem linear to an observer with respect to time. However, if we want to provide a more aesthetically pleasing and realistic animation, an elastic behaviour for the link of the operation would seem appropriate. We can approximate this characteristic by redefining the $f^i$ functions in the above example as $f^i_{x_i,y_i}(t)= x_i + (t-t_s)(t-t_s+1)(y_i-x_i)/((t_e-t_s)(t_e-t_s+1))$ for $1\leq i \leq 8$. This will make the edges move faster as they get closer to each other, providing a "snap" at the end.

We also noted that the problem-solving phase of execution naturally precedes assembly of the final objects. However, this need not be the case. For example, a design may have a structure at the top level similar to the masterkeying network in Figure 13, but the implicit components occurring in it may correspond to designs with a similar structure. For example, the execution of an i-component that is not a literal could produce not only an e-component, but also some structure of function cells.

If the execution of the "problem solving" and construction parts of a network were intermingled, we would need a mechanism for backtracking over assembly steps where operations are applied to solids. Although this may not be problematic for the Lograph subset of LSD, the visual representation of the backtracking needs to reflect this when the program is run in an animated or single step mode so the programmer can step through the assembly of a design specification. This could be done in two ways. The first approach is that whenever a failure occurs, either as a result of two unmergable function cells connected at their head terminals, or a failure is reported to LSD by the solid modeler, the query is dumped and the top-most query from the LSD execution engine replaces it. This approach may be a bit difficult to trace for the user of LSD since the backtracking jumps to a previous state without indicating what operations where undone. The second approach animates operations in reverse order and is a pure reverse playback of the original animation recorded as a movie.

Note that both approaches would require that the solid modeler implement a backtracking mechanism, to be invoked when an operation or an i-component replacement is backtracked over in LSD. This could be achieved through the use of a state stack in the solid modeler. That is, before a new operation is to be executed in the solid modeler, or an i-component is replaced, the state of the solid modeler is stored as a new node on top of the stack. Note that backtracking over deletion, merge, and replacement of literals in LSD would not cause backtracking in the solid modeler.

## 5. Concluding Remarks

We have demonstrated how LSD, a language initially proposed as a means for unifying the design and programming aspects of complex design specifications, can also provide a convenient platform for solving design problems. The visual nature of LSD, combined with the strengths of logic programming that it inherits from Lograph, provide a unique combination for solving such problems.

We also noted that the problem-solving phase of execution naturally precedes assembly of the final objects. However, this need not be the case. The masterkeying example also showed how LSD can allow component level design to be cleanly separated from system level design, where designs describe components and a design specification prescribes the system.

We also discussed the importance of appropriate techniques for editing and debugging LSD programs and mentioned that animation of LSD execution rules may provide valuable feedback to the user of LSD for debugging purposes. While the animation of deletion, merge, and replacement in an implementation of LSD may be inherited from its underlying Lograph implementation, animation of the application of operations to solids require the knowledge of physical properties of solids and therefore must be dealt with by an associated solid modeler. We formally defined animation of operations in solid modeling terms and presented as an example the animation of the bonding operation.

Like programs in other logic programming languages, an LSD program can be run with uninstantiated variables. Similar to fully constrained assemblies, an assembly containing uninstantiated variables stops when the design specification is transformed into an instance of a solid. Note that the uninstantiated variables in a design specification will disappear during the assembly as they are merged with the constants in the program or simply are deleted. For example, imagine a design specification that consists of an implicit component **Key** and a fixed-length list with uninstantiated elements as the input to **Key** (e.g., Figure 6(a) with all the constants except [ ] removed). Assembling such a design specification will produce a key with $n$ bits where $n$ is the length of the input list. Forcing backtracking, in order to find more solutions, will create another key with a different combination of bits. Clearly, by forcing backtracking we can enumerate all possible implementations of an $n$-bit key. However, running a query that consists of an implicit component **Key** with an uninstantiated variable as its input will initially generate a key with no bits (a handle and a tip). Then, by forcing backtracking, we can generate all single-bit keys, two-bit keys, and so forth.

In another thread, LSD can be augmented with constraint logic programming in order to further improve its programming capabilities for practical purposes. We are currently investigating this aspect of LSD and aim to improve the language by integrating recent developments in constraint logic programming.

It is also important to note that LSD is an extendible language in the sense that any operation in any design domain, as defined by Cox and Smedley (2000), can be added to it. Hence LSD can express design problems in any such design domain. Also, given that Lograph is a logic programming language, it has the capacity to represent any computation representable in other logic programming languages such as Prolog.

An implementation of Lograph is presently under way (Banyasad & Cox, 2001 and Banyasad & Cox, 2003). More information on the current status of the Lograph project can be found at `http://www.cs.dal.ca/~vivid/projects/Lograph/lograph.html`.

## Acknowledgements

This research was partially supported by Natural Sciences and Research Council of Canada Discovery Grant OGP0000124.